# Dealing with Sparse Document and Topic Representations: Lab Report for CHiC 2012

Philipp Schaer, Daniel Hienert, Frank Sawitzki,
Andias Wira-Alam, Thomas Lüke

GESIS – Leibniz Institute for the Social Sciences,
Unter Sachsenhausen 6-8, 50667 Cologne, Germany
`firstname.lastname@gesis.org`

**Abstract.** We will report on the participation of GESIS at the first CHiC workshop (Cultural Heritage in CLEF). Being held for the first time, no prior experience with the new data set, a document dump of Europeana with ca. 23 million documents, exists. The most prominent issues that arose from pretests with this test collection were the very unspecific topics and sparse document representations. Only half of the topics (26/50) contained a description and the titles were usually short with just around two words. Therefore we focused on three different term suggestion and query expansion mechanisms to surpass the sparse topical description. We used two methods that build on concept extraction from Wikipedia and on a method that applied co-occurrence statistics on the available Europeana corpus. In the following paper we will present the approaches and preliminary results from their assessments.

## 1  Introduction

In 2011 the CHiC workshop (Cultural Heritage in CLEF) was held for the first time at CLEF. After this initial workshop, where first ideas on how to evaluate the retrieval of "cultural assets" were discussed, CHiC started as a pilot evaluation lab in 2012. Therefore we neither had prior experience with the data set that was provided by the lab organizers nor with the specific domain (cultural heritage) in particular. To allow a systematic IR evaluation the organizers published a dump from the Europeana platform with roughly 23 million documents to form the initial evaluation corpus and a total of 50 different topics that were used in three different tasks. GESIS contributed in two of these tasks: the Ad-hoc Retrieval Task and the Semantic Enrichment Task. In the following paper we will present our approaches and preliminary results from their assessments.

Since this lab had no predecessor we couldn't learn from previous results and best practices. So, the main objective of this initial participation was to establish the retrieval environment, to get a feeling for the data set and to surpass the obvious issues in the first place.

**Table 1.** Word counts for the English Ad-hoc Retrieval Task (ADHOC) and the Semantic Enrichment Task (SE). Given are the mean, median, min and max counts for the description and the title field in the topic file. The ad-hoc used 50 topics (CHIC-001 to CHIC-050) and the semantic enrichment used 25 topics (CHIC-001 to CHIC-025).

| task | field | mean | median | min | max |
| --- | --- | --- | --- | --- | --- |
| ADHOC | description | 2,84 | 0 | 0 | 25 |
| ADHOC | title | 1,94 | 2 | 1 | 6 |
| SE | description | 1,56 | 0 | 0 | 13 |
| SE | title | 1,91 | 2 | 1 | 3 |

The 50 topics provided were very unspecific and underspecified. They consist of the usual identifier, title and description and were provided in three languages (English, French, and German). Only half of them (26/50) contained a description and the titles were usually short with just around two words (see Table 1). As stated by the official lab guideline, the topics "are taken from real Europeana query logs and comprise queries for people, places, work titles (e.g. Mona Lisa), events or subjects". This might explain the sparse representation of the topics, but it never the less is a source of serious retrieval issues that led to the main objectives of our experiments.

When using unprocessed queries on all available metadata fields some topics just produced very small result sets, e.g. topic CHIC-009 ("falkland islands") only returned 12 documents at all. So, we decided to focus on the Semantic Enrichment Task to acquire a rich set of additional query terms that can be used for the necessary query expansion mechanisms in the Ad-hoc Retrieval Task.

In the following paper we will first present our overall technical retrieval system, different filters, and modifiers we used to access the raw data. We will then continue by outlining three different approaches to find appropriate terms for our query expansion. We will conclude with a discussion of the performance of the different approaches and the first lessons learned.

## 2      Indexing and Pre-processing the Europeana Dump

We choose the open-source search framework Solr[1] to index and query the Europeana dump provided by the CHiC organizers. Within Solr it is easy to import the XML data for the evaluation without extensive schema conversions. A main idea behind using Solr in a stand-alone configuration was that the original Europeana platform also uses this technology stack. At GESIS we have made good experiences with Solr in a number of projects like the social science information portal Sowiport[2]. We used the Solr 4 nightly build (build no. 4.0-2012-05-29_08-19-37) for indexing and querying. At the time of the evaluation it was not a stable release but it offered some benefits, like included information analyzers that we used in the evaluation.

---

[1] http://lucene.apache.org/solr/
[2] http://www.gesis.org/sowiport/en/home/overview.html

### 2.1 Solr Configuration

To keep the configuration and schema simple, we used the original Solr configuration and imported the Europeana dump via dynamic fields into the provided schema.xml. Using dynamic fields we stored the information as they are delivered in strings and also for each language in a language based field type. The fieldnames remained as in the original description. For English language information we used the available "text_en" field type, for German language information the "text_de" fieldtype.

The available language specific field types in Solr 4 offer the following analyzers[3] for all languages:

- StandardTokenizerFactory: A general purpose tokenizer, which divides a string into tokens with various types.
- StopFilterFactory: Words from the Solr included stopword lists are discarded.
- LowerCaseFilterFactory: All letters are indexed and queried as lowercase.

Additionally, language specific analyzers[4] were used, for English and German:

- EnglishPossessiveFilterFactory,
- PorterStemFilterFactory: A stemmer for English,
- GermanNormalizationFilterFactory,
- GermanLightStemFilterFactory: A stemmer for German.

With the use of a copyField we stored all separate field information in a common search field (chic-all).

### 2.2 Frequencies of Europeana Metadata Fields

After indexing the Europeana information in Solr, we were able to create an overview about the coverage of different fields for English and German (see Table 2). We see that the "europeana" namespace is nearly completely available for all datasets. The basic namespace, that includes many available information is the Dublin Core ("dc") namespace. After that follows the "enrichment" namespace and rarely filled are fields from the "dcterms" namespace. In our evaluation we will concentrate our queries to data fields that are mostly available.

## 3 Acquiring Related Concepts

We introduce three different techniques to acquire related concepts that would later be used to allow a query expansion. We tried to find related concepts in Wikipedia out-links from the lead section (Section 3.1), from the Wikipedia full text using document similaritiy (Section 3.2) and the Europeana corpus itself using co-occurrence analyses (Section 3.3).

---

[3] https://wiki.apache.org/solr/AnalyzersTokenizersTokenFilters
[4] https://wiki.apache.org/solr/LanguageAnalysis

**Table 2.** Overview on the coverage of data fields in the Europeana dump used for CHIC. The second and fourth columns indicate the percentage coverage of each field. So, in 14% of all documents the dc:contributor field contained some data, while the europeana:country field was filled in every document.

| Field name | % | Field name (continued) | % |
| --- | --- | --- | --- |
| dc:contributor | 14 | dcterms:tableOfContents | 0 |
| dc:coverage | 30 | dcterms:temporal | 4 |
| dc:creator | 57 | enrichment:agent_label | 0 |
| dc:date | 50 | enrichment:agent_term | 0 |
| dc:description | 61 | enrichment:concept_broader_label | 35 |
| dc:format | 56 | enrichment:concept_broader_term | 35 |
| dc:identifier | 98 | enrichment:concept_label | 37 |
| dc:language | 42 | enrichment:concept_term | 37 |
| dc:publisher | 34 | enrichment:period_broader_label | 45 |
| dc:relation | 43 | enrichment:period_broader_term | 45 |
| dc:rights | 62 | enrichment:period_label | 45 |
| dc:source | 68 | enrichment:period_term | 45 |
| dc:subject | 50 | enrichment:place_broader_label | 10 |
| dc:title | 98 | enrichment:place_broader_term | 10 |
| dc:type | 88 | enrichment:place_label | 13 |
| dcterms:alternative | 3 | enrichment:place_term | 13 |
| dcterms:created | 22 | europeana:country | 100 |
| dcterms:extent | 22 | europeana:dataProvider | 78 |
| dcterms:hasFormat | 2 | europeana:isShownAt | 99 |
| dcterms:hasPart | 0 | europeana:isShownBy | 51 |
| dcterms:hasVersion | 16 | europeana:language | 100 |
| dcterms:isPartOf | 15 | europeana:object | 97 |
| dcterms:isReferencedBy | 0 | europeana:provider | 100 |
| dcterms:issued | 1 | europeana:rights | 60 |
| dcterms:medium | 13 | europeana:type | 100 |
| dcterms:provenance | 7 | europeana:uri | 100 |
| dcterms:references | 0 | europeana:year | 47 |
| dcterms:spatia | 0 | | |

### 3.1 Extracting Concepts from Wikipedia Lead Sections

In this approach we use related concepts from Wikipedia summaries to semantically enrich CHIC topics. Wikipedia articles represent important concepts of the world knowledge. If we can find a Wikipedia article that represents the topic, we can use the text of the article to extract important concepts that are related to it. In this implementation we used links from the lead section, which summarizes the whole article with the most important aspects. Wikipedia guidelines for the lead section[5] suggest: "The lead should be able to stand alone as a concise overview. It should define the topic, establish context, explain why the topic is notable, and summarize the most important points […]". Therefore it could be well-suited to find nearby related concepts. Links in the summary represent relations to other important concepts existing in Wikipedia, which we use for the semantic enrichment.

The approach therefore consists of two steps: (1) finding an appropriate Wikipedia article that represents the topic and (2) extracting links from the article's first paragraph as a representation of important concepts.

For the first step, we have created a SOLR index of all titles of Wikipedia articles. In several iterations we search for the topic with the (a) original topic, (b) the topic excluding stop words, (c) the permutation of topic words and (d) individual words from the topic. The title with the highest TF*IDF score is then used as a representation. For nearly all topics we were able to find a Wikipedia article that represents it. Problems occur with concepts not contained in Wikipedia (like topic CHIC-049 "teufelstal"), very broad topics (like topic CHIC-020 "europa maps 1914") or topics that must be searched not only by the title, but in the full text (topic CHIC-037 "1809 combat")

In the next step we extracted all links from the summary of the article. Therefore, we first took the original WikiSyntax from the article through the Wikipedia API. Then, we clean the article text from internal/system links and other sections like info boxes. As a next step, we extract the summary above the first header. Links from this section are then extracted with regular expressions, utilizing the fact that they are marked with double square brackets. If we do not get enough links from this, we use the whole article text, for example for very short articles. The extracted links are then used as semantic enrichments for the original topic.

### 3.2 Extracting Concepts from Wikipedia Full Texts using Document Similarity

In this second approach we use Wikipedia full texts to extract related concepts. For our training corpus we use a particular subset of Wikipedia. In contrast to the previous approach we map Wikipedia entries of the given terms manually.

We used two different Wikipedia training corpora to enrich the query terms. The first corpus is the German Wikipedia corpus that consists of 1,054,842 articles. The articles are here randomly selected and the corpus comprises almost the half of the complete German Wikipedia corpus[6]. The second corpus contains only articles having backlinks or outlinks to the particular English Wikipedia entries taken from the 50 CHIC query terms. We extracted the back- and outlinks using the RelExAPI tool[7] (also used [5]). This is based on the assumption that all articles linked to the source article are somehow related, therefore it will help later to find the related terms. Overall, we have extracted 85,847 articles of the English Wikipedia. Finally, we crawl each article page, extract the page contents, strip the HTML tags, and store them into individual text files for the training corpus.

---
[5] http://en.wikipedia.org/wiki/Wikipedia:Manual_of_Style_%28lead_section%29

As an initial step, we index each training corpus using Lucene[8]. Since each term is represented as an article, given a query term, the related terms are the articles that are most similar to the source. Therefore, this task can also be seen as finding related documents. Formally, we define the similarity score of two documents, denoted as $d_1$ and $d_2$, as follows

$$sim(d_1, d_2) = \frac{1}{n} |d_1 \cap d_2|$$

where $d_1$ and $d_2$ are vectors of words with cardinality of $n$. We set a factor $n$ to denote the number of important words included in the calculation. The importance of a word can be obtained by calculating its TF*IDF score. This method is a slight modification of the Jaccard similarity coefficient.

### 3.3  Extracting Concepts from Europeana using Co-occurrence Analyses

A common approach to find related concepts is the use of co-occurrence analysis. It is presented extensively in standard natural language processing handbooks e.g. the one by Manning and Schütze [2]. Co-occurring elements are such elements that are likely to occur in the same context. To increase retrieval performance we extracted those terms from the dataset that are most likely to co-occur with the terms of a given topic. This approach was implemented and used by us in other query expansion scenarios [1, 4] and the general idea was presented as the so-called Search Term Recommender by Petras [3], although the original concept based on controlled vocabularies.

In order to specify which terms are likely to co-occur a similarity measure has to be defined. In our approach we choose the Jaccard Index which is calculated for two attributes x and y where $DS_x$ and $DS_y$ are two document sets with $DS_x$ containing documents with attribute x and $DS_y$ with attribute y, respectively. $DS_{xy}$ is the documents set containing attributes x and y (i.e. the intersection of $DS_x$ and $DS_y$). The document frequencies $df_x$, $df_y$ and the collocation frequency $df_{xy}$ are defined as the size of $DS_x$, $DS_y$ and $DS_{xy}$. The Jaccard similarity is given in the next equation.

$$J(x,y) = \frac{|DS_x \cap DS_y|}{|DS_x \cup DS_y|} = \frac{df_{xy}}{df_x + df_y - df_{xy}}$$

---

[6] As of June 3, 2012, the German Wikipedia had 2,398,859 articles.
[7] http://multiweb.gesis.org/RelExAPI
[8] http://lucene.apache.org/

To cope with large differences in the size of $DS_x$ and $DS_y$ we modified the index by taking the logarithm.

With this measure we processed the following attributes from the Europeana corpus. Terms appearing in dc:title and/or dc:description were treated as input (i.e. query terms) and their co-occurrence with terms from the fields dc:subject and enrichment:concept_label was measured using the Jaccard Index. To make the process easier to understand let us look at the following example. Topic CHIC-010 consists of the query "film canada". Therefore, we measured which concepts from dc:subject and enrichment:concept_label co-occurred the most with the terms "film" and "Canada" in title or description. The resulting top 3 concepts are: "poster", "Cinema and Theatre", and "popular media". These concepts appear to be semantically related to the query. However, it also becomes clear that the quality of the related concepts can only be as good as the quality of the vocabulary used in dc:subject and enrichment:concept_label.

## 4 Result Set Construction for the Ad-hoc and Semantic Enrichment task

Each method listed in Section 3 returned a list of ten concepts that we used as the result sets for the Semantic Enrichment Task and to establish a rudimentary query expansion mechanism for the Ad-hoc Retrieval Task. Using the previously described methods we established the following concept extraction services (the abbreviations are the same used for the names of the runs and in the DIRECT system):

- WIKI_ENTITY – Concepts from the Wikipedia lead section, extracted by detecting outlinks (Section 3.1).
- WIKI_SIM – Concepts from the Wikipedia full text, extracted from the German Wikipedia subset (1 million documents) using document similaritiy (Section 3.2).
- WIKI_BACK – Concepts from the Wikipedia full text, extracted from a Wikipedia subset using document similarities. This subset consists of all back- and outlinks of a given seed document (Section 3.2).
- STR – Concepts from the Europeana data set, extracted using co-occurrence analyses. The analyses were language dependent, so only the specific language corpus was used (Section 3.3).
- COMBO – A mixture of all available concepts from the four previous services

The result sets for the Semantic Enrichment task were constructed from the top 10 ranked results from the previously described approaches. While the co-occurrence and Wikipedia document similarity approaches returned a ranked list based on the similarity scores, the Wikipedia concept extraction from the lead sections did not include such a score. We used the implicit ranking due to order of their appearance in the text. We suppose that the earlier a linked concept appears in the summary the more important is has to be.

The query construction for the Ad-hoc Retrieval Task was done by taking the original title, removing stopwords and adding the concepts by OR-ing them with the title terms. We boosted the title term by factor 2 (^2) and searched in all available metadata fields in the current language (chic-all, see Section 2). The expanded query for topic CHIC-012 ("moby dick") and the concepts gained from the Wikipedia lead sections therefore looks like the following example:

```
chic_all-en:(moby OR dick)^2 OR chic_all-en:("Herman Mel-
ville" OR "English language" OR "Adventure novel" OR "Sea
story" OR "Richard Bentley" OR "Harper Brothers" OR "Her-
man Melville" OR "The Great American Novel" OR "litera-
ture" OR "Ishmael (Moby-Dick)")
```

Table 3. Overview on the different methods that were used for the two monolingual Ad-hoc tasks and the two monolingual Semantic Enrichment tasks. The WIKI_SIM system was only used for the first 25 topics in the ADHOC-DE-DE task, since the other topics couldn't be generated by the system on time.

|  | ADHOC-EN-EN | ADHOC-DE-DE | SE-EN-EN | SE-DE-DE |
| --- | --- | --- | --- | --- |
| WIKI_ENTITY | ✓ | ✓ | ✓ | ✓ |
| WIKI_SIM | - | ✓/- | ✓ | ✓ |
| WIKI_BACK | ✓ | - | ✓ | - |
| STR | ✓ | ✓ | ✓ | ✓ |
| COMBO | ✓ | ✓ | - | - |

The systems and concepts used for the expansion are identical to the ones from the Semantic Enrichment Task. Additionally a combination of all concepts was submitted (combo). The different combinations and systems used for each tasks and the results we submitted for the Ad-hoc Retrieval and the Semantic Enrichment Task are listed in Table 3. All queries are stemmed and pre-processed at query time by the Solr filters that are listed and described in Section 2.

## 5 Results

In the following section we report on the results from the different implementations presented in the previous sections. As we participated in two tasks we will present the results according to each task. A summary of the results is given in Table 4 and 5. Due to a problem in the DIRECT evaluation system we could not access all data. The figures are based on our own calculations using trec_eval, while the tables are based on the data included in the figures provided by DIRECT.

### 5.1 Ad-hoc Retrieval Task

Out of the eight system runs we submitted the WIKI implementations could generally produce the most effective results that were above the average performance of all competing systems (AVG_ALL) which was 0.4255 for the English sub-task and 0.5111 for the German. The WIKI_ENTITY system was the best among our systems with MAP value of 0.4396 (EN) and 0.5680 (DE). In both cases the WIKI_ENTITY was among the top 5 participating systems.

In the English sub-task 14-15 topics didn't returned any documents which resulted in an empty result set and a MAP value of 0. In the German sub-task only 4-6 topics had this problem. Some other topics nearly produced the same MAP value (like CHIC-001, CHIC-006, CHIC-014 or CHIC-034) which might be seen as an indicator that the terms used to expand the query didn't have any effect.

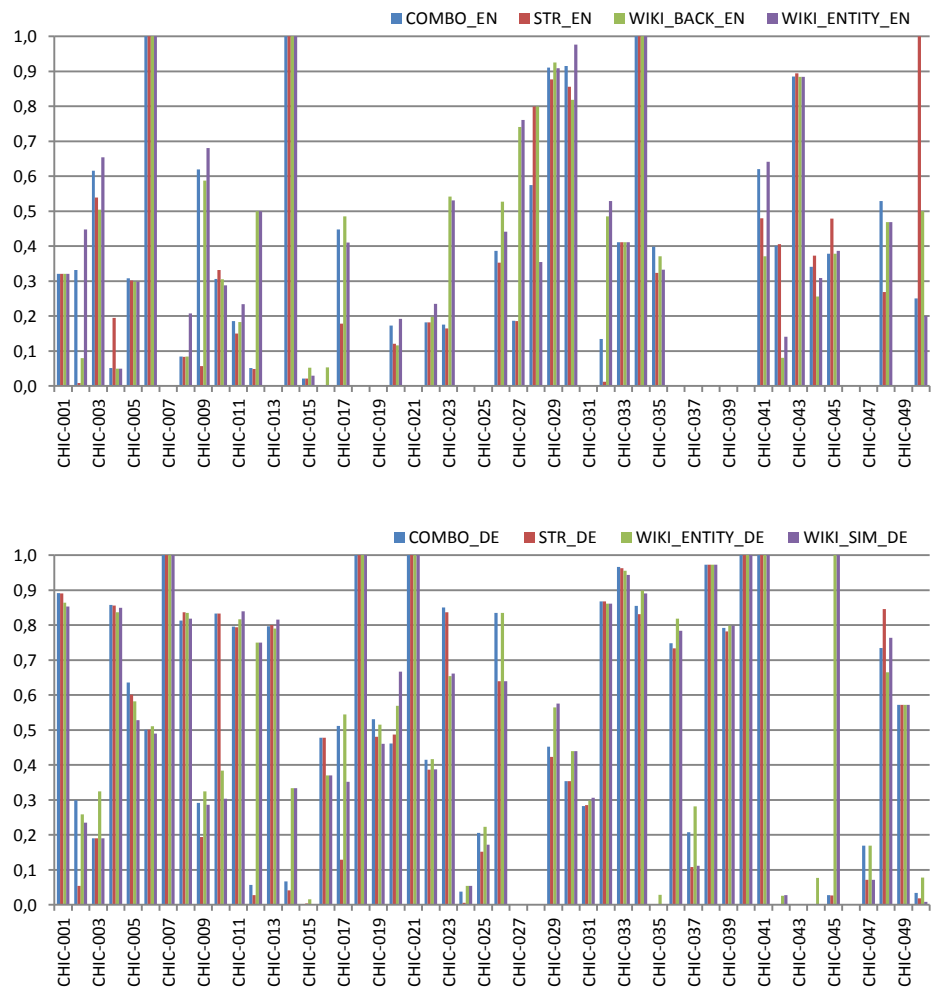

**Fig. 1.** Plot of MAP values for the English and German monolingual Ad-hoc Retrieval Task.

**Table 4.** Results of the assessment for the Ad-hoc Task. We report on mean average precision (MAP) and R-Precision (R-P) of each system for each language. Additionally the average values of all systems that participated in the task is reported (AVG_ALL).

| System | MAP (EN) | R-P (EN) | MAP (DE) | R-P (DE) |
|---|---|---|---|---|
| WIKI_ENTITY | 0.4396 | 0.4380 | 0.5680 | 0.5362 |
| WIKI_BACK | 0.4271 | 0.4116 | - | - |
| WIKI_SIM | - | - | 0.5246 | 0.5190 |
| STR | 0.3728 | 0.3625 | 0.4806 | 0.4764 |
| COMBO | 0.3944 | 0.3887 | 0.5081 | 0.5040 |
| AVG_ALL | 0.4255 | 0.4175 | 0.5111 | 0.5036 |

**Table 5.** Results of the assessment for the Semantic Enrichment Task for the systems WIKI_ENTITY, WIKI_BACK, WIKI_SIM, and STR. Additionally the average precision of all systems that participated in the task is reported (AVG_ALL). Strong precision (P-strong) is the average precision (over 25 queries) of "relevant" suggestions over all suggestions. Weak precision (P-weak) is the average precision (over 25 queries) of "relevant" and "maybe relevant" over all suggestions.

| System | P-weak (EN) | P- strong (EN) | P-weak (DE) | P strong (DE) |
|---|---|---|---|---|
| WIKI_ENTITY | 0.9240 | 0.7000 | 0.8794 | 0.7448 |
| WIKI_BACK | 0.6440 | 0.5200 | - | - |
| WIKI_SIM | 0.6320 | 0.5080 | 0.1160 | 0.0840 |
| STR | 0.1800 | 0.1000 | 0.1760 | 0.0960 |
| AVG_ALL | 0.6834 | 0.5470 | 0.6045 | 0.4721 |

### 5.2 Semantic Enrichment Task

Out of the seven runs that we submitted to the Semantic Enrichment task, the WIKI_ENTITY implementation could outperform both our own and the competing implementations from other groups. In the manual assessment this approach could achieve a precision of 0.9240 (weak)/0.7000 (strong) in the English monolingual run, and 0.8794 (weak), 0.7448 (strong) in the German monolingual run (see Table 5). All other implementations were below the average precision over all runs. While for the English run the other both WIKI systems could achieve precision values that were only slightly below the average, the STR system could only provide useful enrichments in 1/5 (weak) or 1/10 (strong) of the cases.

When evaluated as a Query Expansion mechanism by the CHiC organizers, the WIKI_ENTITIY system was still among the top 5 systems with MAP values of 0.2338 (EN) and 0.3192 (DE). Surprisingly the reference implementation ORIGINALQUERIESEN by the organizers was the best systems with MAP of 0.3411 (EN) and 0.5701 (DE). Our system outperformed the others in topics CHIC-

001 and CHIC-017, while it was significantly worse in CHIC-005. Since the implementation details are not clear[9], we cannot describe this any further.

## 6 Discussion

In this paper we described three different approaches that were implemented in five different systems (see Table 3).

For both Wikipedia methods the mapping of the topic title to a Wikipedia document were an essential first step. We surpassed this by using a rather ad-hoc implementation using a separate Solr index (WIKI_ENTITY) or by mapping them by hand (WIKI_SIM and WIKI_BACK). We are aware of the fact that there are public APIs to access the Wikipedia content in a more convenient way, like JWPL [6], but for more flexibility in adjusting details we choose to implement these routines ourselves. In a future version, we also want to extract entities from the article's first paragraph or the whole article that are not marked as links. Many concepts in the free text have not been linked by users to their Wikipedia articles or the article is still missing. This could further improve the semantic enrichment of topics with nearby concepts.

The performance of the STR was worst in all tasks and did not perform very well compared with most other approaches. A possible explanation for this is the lack of consistent controlled vocabulary in Europeana documents and the fact that in at least 50% of all cases no entry was made (see Table 2) in the fields used for co-occurrence analysis (dc:subject and enrichment:concept_label). Looking at the concepts suggested by our service one might also doubt the consistency of the vocabulary used in Europeana. Topic CHIC-019 might serve as an example. The topics title is *philosophical anthropology* and the resulting concepts suggested by our STR service are: *звук, sunet, ήχος, sonido, dźwięk, garsas, zvuk, lyd, ääni,* and *suono.* While the meaning of these recommendations is not always clear it is apparent that they are of the wrong language. These were retrieved from the English dataset and should thus be of English language.

These inconsistencies are also reflected in the results of the assessment campaign. While our approach does perform better than some others it is also clearly surpassed by approaches which use more controlled vocabulary (e.g. our Wikipedia entities approach). Extracting concepts using co-occurrence analysis works well if the given dataset is of high and consistent quality and uses a controlled vocabulary on the majority of its entries. However, in the given case of Europeana approaches that make use of external knowledge are better suited.

The WIKI_BACK system in the Ad-hoc task could produce quite satisfying results that were just under the WIKI_ENTITY system. For the query terms that have Wikipedia entries, this method provides in some cases reasonable results. In contrast, if the query terms are very general or do not exist as Wikipedia entries, this method would provide "erratic" results. The query terms, e.g. "zeppelin 1900", "england cup final", "unarmed", or "europe maps 1914", are difficult to be mapped into Wikipedia entries. We have no prior knowledge of what users are actually looking for and we cannot simply adjust these terms into particular Wikipedia entries. As a consequent, we deal with uncertainty in this matter in order to enrich the query terms with Wikipedia and therefore the results provided by this method are rather ambiguous.

---

[9] "[…] we used a Lucene index to compare runs with just the original queries (those runs are marked with ORIGINALQUERIES in the title) to runs that included the original queries plus the semantic enrichment suggested by your experiments. These runs were now assessed for relevance exactly like the ad-hoc runs and you can compare your results with all the standard metrics." (excerpt from an organizer's email)

In general we could establish a retrieval environment that while being technically comparable with the Europeana (Solr-based) system, surpasses some of the previously described problem in respect to the sparse document and topic representations. Never the less a lot of issues remain unsolved.

**Acknowledgements.** This work was partially supported by Deutsche Forschungsgemeinschaft (DFG, German Research Foundation), grant number SU 647/5-2.